\documentclass[sigconf,screen]{acmart}

 \setcopyright{none}
 \renewcommand\footnotetextcopyrightpermission[1]{} 

\newcommand{{\methodname}}{ROME}
\usepackage{balance}
\usepackage{multirow}
\usepackage{algorithm}
\usepackage{tcolorbox}
\usepackage[noend]{algpseudocode}
\usepackage{pgfplots} 
\pgfplotsset{compat=1.8}
\usepackage{pgf-pie} 
\usepackage{url}

\RequirePackage[normalem]{ulem} 
\RequirePackage{color}\definecolor{RED}{rgb}{1,0,0}\definecolor{BLUE}{rgb}{0,0,1} 
\providecommand{\DIFdelbegin}{} 
\providecommand{\DIFdelend}{} 
\providecommand{\DIFaddbeginFL}{} 
\providecommand{\DIFaddendFL}{} 
\providecommand{\DIFdelbeginFL}{} 
\providecommand{\DIFdelendFL}{} 

\usepackage{amsmath,amsfonts}
\usepackage{graphicx}
\usepackage{textcomp}
\usepackage{pbox}
\usepackage{xcolor}

\AtBeginDocument{%
  \providecommand\BibTeX{{%
    \normalfont B\kern-0.5em{\scshape i\kern-0.25em b}\kern-0.8em\TeX}}}

\setcopyright{acmlicensed}
\acmPrice{15.00}
\acmDOI{10.1145/3597926.3598094}
\acmYear{2023}
\copyrightyear{2023}
\acmSubmissionID{issta23main-p216-p}
\acmISBN{979-8-4007-0221-1/23/07}
\acmConference[ISSTA '23]{Proceedings of the 32nd ACM SIGSOFT International Symposium on Software Testing and Analysis}{July 17--21, 2023}{Seattle, WA, USA}
\acmBooktitle{Proceedings of the 32nd ACM SIGSOFT International Symposium on Software Testing and Analysis (ISSTA '23), July 17--21, 2023, Seattle, WA, USA}

\begin{document}

\title{ROME: Testing Image Captioning Systems via Recursive Object Melting}
\author{Boxi Yu}
\email{boxiyu@link.cuhk.edu.cn}
\affiliation{
  \institution{The Chinese University of Hong Kong, Shenzhen}
  \city{Shenzhen}
  \state{Guangdong}
  \country{China}
}

\author{Zhiqing Zhong}
\email{cheehingchung@gmail.com}
\affiliation{%
  \institution{The Chinese University of Hong Kong, Shenzhen}
  \city{Shenzhen}
  \state{Guangdong}
  \country{China}
}

\author{Jiaqi Li}
\email{jiaqili3@link.cuhk.edu.cn}
\affiliation{%
  \institution{The Chinese University of Hong Kong, Shenzhen}
  \city{Shenzhen}
  \state{Guangdong}
  \country{China}
}

\author{Yixing Yang}
\email{120090268@link.cuhk.edu.cn}
\affiliation{%
  \institution{The Chinese University of Hong Kong, Shenzhen}
  \city{Shenzhen}
  \state{Guangdong}
  \country{China}
}

\author{Shilin He}
\email{shilin.he@microsoft.com}
\affiliation{%
  \institution{Microsoft Research}
  \city{Beijing}
  \country{China}
}

\author{Pinjia He}
\authornote{Corresponding author.}
\email{hepinjia@cuhk.edu.cn}
\affiliation{%
  \institution{The Chinese University of Hong Kong, Shenzhen}
  \city{Shenzhen}
  \state{Guangdong}
  \country{China}
}


%



\keywords{Metamorphic testing, testing, image captioning, AI software}



\begin{abstract}
Image captioning (IC) systems aim to generate a text description of the salient objects in an image.
In recent years, IC systems have been increasingly integrated into our daily lives, such as assistance for visually-impaired people and description generation in Microsoft Powerpoint.
However, even the cutting-edge IC systems (\textit{e.g.}, Microsoft Azure Cognitive Services) and algorithms (\textit{e.g.}, OFA) could produce erroneous captions, leading to incorrect captioning of important objects, misunderstanding, and threats to personal safety.
The existing testing approaches either fail to handle the complex form of IC system output (\textit{i.e.}, sentences in natural language) or generate unnatural images as test cases. 
To address these problems, we introduce \textit{Recursive Object MElting} ({\methodname}), a novel metamorphic testing approach for validating IC systems. 
Different from existing approaches that generate test cases by inserting objects, which easily make the generated images unnatural, {\methodname} \textit{melts} (\textit{i.e.}, remove and inpaint) objects.
{\methodname} assumes that the object set in the caption of an image includes the object set in the caption of a generated image after object melting.
Given an image, {\methodname} can recursively remove its objects to generate different pairs of images.
We use {\methodname} to test one widely-adopted image captioning API and four state-of-the-art (SOTA) algorithms.
The results show that the test cases generated by {\methodname} look much more natural than the SOTA IC testing approach and they achieve comparable naturalness to the original images.
Meanwhile, by generating test pairs using 226 seed images, {\methodname} reports a total of 9,121 erroneous issues with high precision (86.47\%-92.17\%).
In addition, we further utilize the test cases generated by {\methodname} to retrain the Oscar, which improves its performance across multiple evaluation metrics.

\end{abstract}

\maketitle

\section{Introduction}\label{sec:intro}

Image captioning (IC) systems, which generate a text description of the salient objects given an image, have been widely adopted in real-life scenarios.
For instance, the geographic information system (GIS) ArcGIS~\cite{arcgis_api} employs IC to describe the salient objects in remote sensing images for accessibility purposes.
Microsoft Edge Explorer uses IC to help visually-impaired individuals understand images on the web~\cite{edge_api}.
Meta Facebook utilizes IC to provide alternative texts for user photos~\cite{facebook_api}.  
These recent advances have been largely facilitated by the rapid development of deep neural networks in both computer vision (CV) and natural language processing (NLP)~\cite{vinyals_show_2015, xu_show_2015, li_oscar_2020, zhang_vinvl_2021, hu_vivo_2020, wang2022unifying}.
In particular, recent research~\cite{hu_vivo_2020} has claimed that the SOTA IC algorithm has achieved human-level performance in terms of CIDEr score~\cite{vedantam2015cider}.

\begin{figure}[ht]
		\centering
		\DIFdelbeginFL 
\DIFdelendFL \DIFaddbeginFL \includegraphics[width=0.83\linewidth]{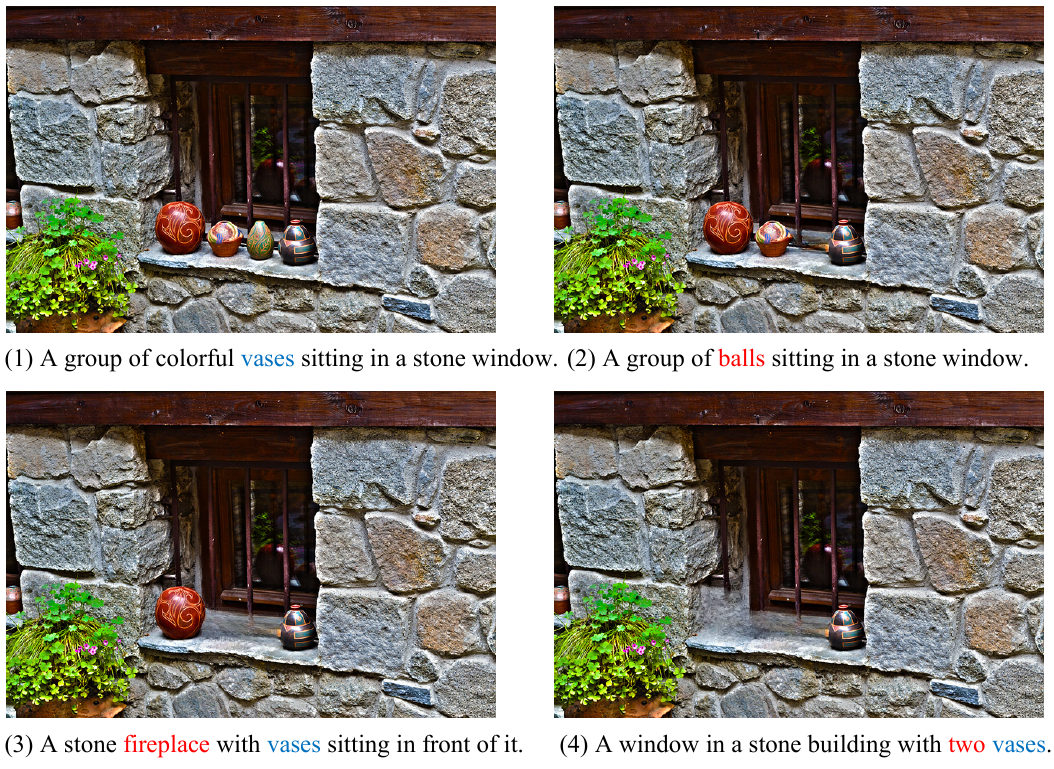}
		\DIFaddendFL \caption{The original image (1) and three images (2-4) generated by {\methodname} and the corresponding captions returned by VinVL~\cite{zhang_vinvl_2021}. Text in blue indicates the salient object(s) in the image, and text in red indicates caption error.}
		\label{fig:rome_recur_show}
		\vspace{-1em}
\end{figure}

Nevertheless, even top-notch IC systems are susceptible to generating incorrect captions, which can lead to severe misunderstandings, negative social consequences, or threats to personal safety~\cite{news_1_ai_nodate,ic2impariment,schroeder_microsoft_2016}.
Fig.~\ref{fig:rome_recur_show} presents four images and the corresponding captions generated by a recent IC algorithm~\cite{zhang_vinvl_2021}.
We can observe that the caption for image (4) "\textit{A window in a stone building with \underline{two} vases}") is incorrect since there is only one vase in the image.
According to media reports, the captions produced by current IC systems are often inaccurate or overly generic (\textit{e.g.}, incorrect gender or nationality)~\cite{ai_alttext_terrible}.
Additionally, a recent report by World Health Organization~\cite{organization_world_2019} shows that more than 2.2 billion people have vision impairment and they are expected to benefit from IC systems-powered AI assistants~\cite{ic2impariment,ic2impariment_2}.
However, since these IC systems can generate incorrect captions, AI assistants may return misleading messages, putting visually-impaired users at risk \cite{news_1_ai_nodate}.
Therefore, it is crucial to ensure the resilience and reliability of modern IC systems.

Developing an effective testing approach is a critical step towards achieving a robust image captioning (IC) system, similar to traditional software.
However, it is a challenging task due to several reasons.
Unlike traditional software, where the logic is predominantly expressed in the source code, the logic of IC systems is primarily encoded in the underlying neural network.
Consequently, code-based testing approaches~\cite{romdhana2021cosmo, viticchie2016assessment} designed for traditional software are not suitable for testing IC systems.
Furthermore, existing testing techniques for AI software mainly focus on software with a simple output format (\textit{e.g.}, a label for an image classifier).
In contrast, the output of an IC system is a natural language sentence, which is significantly more complex, rendering the ineffectiveness of existing approaches (\textit{e.g.}, metamorphic relations that assume the output label should be the same after perturbation).
The SOTA image captioning testing approach MetaIC~\cite{yu2022automated} achieved high precision in reporting suspicious issues.
In particular, MetaIC generates test cases by inserting an object into a specific position in a background image.
Due to the difficulty of finding a suitable insertion position, MetaIC often generates unnatural test cases (\textit{e.g.}, an image with "\textit{flying monkey around the moon}"), which are less likely to occur in the real world.

To tackle these challenges, we introduce \textit{recursive object melting} ({\methodname}), a novel, wide-applicable methodology for validating image captioning systems.
Different from the SOTA approaches that rely on object insertion to generate test cases, which can easily insert an object into an unsuitable position or with an unnatural size (\textit{e.g.}, a cat is larger than a plane), {\methodname} \textit{melt} an object.
Specifically, to melt an object, {\methodname} removes an object from an image and then conducts image inpainting (\textit{i.e.}, recover the missing part based on the remaining context of the image).
Our core idea is that the object set in the caption of an image should include the object set in the caption of a generated image after melting.
For example, in Fig.~\ref{fig:rome_recur_show}, the original image (\textit{i.e.}, image (1)) has four vases in a window.
When one vase was melted, we obtained the second image.
Intuitively, the objects in the caption of this generated image should be included in that of the original image.
However, we observed that object "balls" in the caption "\textit{A group of \underline{balls} sitting in a stone window}" was not included in the object set of the original image, indicating a potential error in either the caption of the original image or that of the generated image.
Based on this idea, {\methodname} recursively melts the object in an image (\textit{e.g.}, the other three images in Fig.~\ref{fig:rome_recur_show}) and constructs multiple image pairs.
Accordingly, we realize {\methodname} as two metamorphic relations (MRs): (1) \textit{object melt}: the object set of a melted image is the subset of that of the image before melting; (2) \textit{class melt}: an object class should not be in the caption if all the objects in that class were melted.
Specifically, {\methodname} automatically melts images by LaMa~\cite{suvorov2022resolution}, an advanced image inpainting technique, and identifies the object names from the captions by Stanza's POS Tagging technique~\cite{stanza}.
If the captions of the image pair returned by the IC system violate any of these MRs, the image pair and the corresponding captions will be reported as a suspicious issue.

To evaluate the effectiveness of {\methodname}, we use it to test one commercial API (\textit{i.e.}, Microsoft Azure Cognitive Services) and four SOTA IC algorithms (\textit{i.e.}, OFA~\cite{wang2022unifying}, Oscar~\cite{li_oscar_2020}, VinVL~\cite{zhang_vinvl_2021}, Attention~\cite{xu_show_2015}).
Based on a user study, test cases generated by {\methodname} are significantly more natural compared to those produced by the SOTA IC testing algorithm~\cite{yu2022automated}.
In the user study, annotators rated the naturalness of the test cases on a scale from 1 (unnatural) to 4 (natural), and the naturalness ratings were 3.33 \textit{vs.} 1.41 and 3.33 \textit{vs.} 1.37 for two running modes of the SOTA IC testing algorithm MetaIC~\cite{yu2022automated}.
{\methodname}'s test cases were even reported to be almost as natural as the original images (3.33 \textit{vs.} 3.56).
By generating test pairs from 226 seed images, {\methodname} reports 9,121 erroneous issues with high precision (86.47\%-92.17\%), which is comparable to the SOTA approach.
Meanwhile, we adopt {\methodname} to unveil labeling errors of COCO Caption~\cite{chen2015microsoft}, and successfully report 219 labeling errors.
In addition, we leverage the test cases generated by {\methodname} to retrain the Oscar base model, which improves its performance in terms of BLEU-4 (1.5\%), METEOR (3.4\%), CIDEr (2.0\%), and SPICE (7.9\%). The source code and the dataset of this paper will be publicly released.

This paper makes the following main contributions.
\DIFdelbegin 

\DIFdelend \begin{itemize}
    \item The introduction of a novel, widely-applicable methodology, recursive object melting ({\methodname}), for image captioning software validation;
    \item A practical implementation of {\methodname} by using LaMa~\cite{suvorov2022resolution} for image inpainting and leveraging Stanza's POS Tagging~\cite{stanza} for object name identification. 
    \item Empirical results demonstrating the effectiveness of {\methodname}: {\methodname} achieves comparable precision (86.47\% $\sim$92.17\%) to the SOTA approach~\cite{yu2022automated}, and the test cases generate by {\methodname} are much more natural (3.33 \textit{vs.} 1.26) in terms of human-annotated naturalness score.
\end{itemize}
 \DIFdelbegin 

\DIFdelend

\section{Preliminaries}\label{sec:bg}

\subsection{Image Captioning Algorithms}
Typically, IC is an image-to-sequence problem, where the input is image pixels and the output is natural language sentences. 
Previously, IC algorithms mainly entailed description retrieval~\cite{pan2004automatic, farhadi2010every,ordonez2011im2text} and hand-crafted natural language generation approaches~\cite{aker2010generating, li2011composing, yang2011corpus}.
With the development of deep learning, modern IC algorithms contain two steps: (1) \textit{visual encoding}: encode the input as one or multiple feature vectors; (2) \textit{sentence generation}: decode the intermediate representations into a sequence of words or sub-words according to a vocabulary. There are mainly two lines of IC systems: LSTM-based IC and multimodal pretraining IC. For LSTM-based IC, researchers proposed the single-layer LSTM~\cite{vinyals_show_2015}, two-layer LSTM~\cite{donahue2015long}, and LSTM with additive attention graph~\cite{xu_show_2015}.

In this paper, we mainly test multimodal pretraining IC algorithms, which have been developed rapidly in recent years~\cite{lu2019vilbert, chen2020uniter, gan2020large}.
Oscar~\cite{li_oscar_2020} adopts a BERT-like architecture that utilizes object tags as anchors to ease the semantic alignment between images and texts. 
Based on Oscar, VinVL~\cite{zhang_vinvl_2021} enhances feature extraction by adopting a specifically designed object detection component. OFA~\cite{wang2022unifying} achieves SOTA performance by unifying several vision and language tasks via a sequence-to-sequence learning framework, which has a unified instruction-based task representation.

\subsection{Motivating Examples}
\label{subsec:motivating_example}
\begin{figure}[t]
		\centering
		\includegraphics[width=0.98\linewidth]{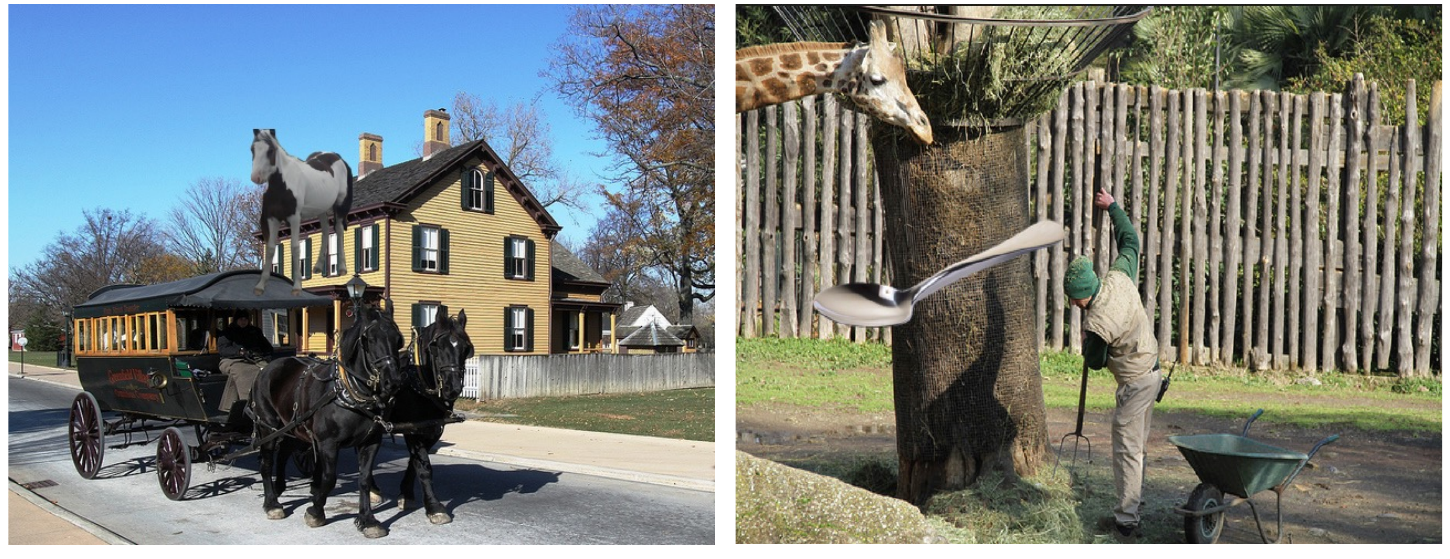}
		\caption{A motivating example of test cases generated by MetaIC.}
		\label{fig:metaic_failure}
\end{figure}

Most of the SOTA testing approaches for IC and its related tasks generate test cases by inserting an object into a seed image. 
In particular, MetaOD~\cite{wang2020metamorphic}, which is the SOTA testing approach for object detection, inserts an object to a location that does not overlap with existing objects in the seed image. 
MetaIC~\cite{yu2022automated}, the SOTA testing approach for IC systems, inserts an object into a background image after location tuning and object resizing. 
Although these approaches perform well in finding erroneous outputs, the test cases generated by them are often unnatural.
For example, Fig.~\ref{fig:metaic_failure} demonstrates two test cases generated by MetaIC, where it inserts a horse into the image on the left and a spoon into the image on the right.
We can observe that even MetaIC provides location tuning and object resizing strategies, the images are still unnatural.
The left-hand side image inserts a horse into an inappropriate location (\textit{i.e.}, on the roof), while the right-hand side image inserts an oversized spoon, making it larger than a human.

In this paper, naturalness refers to how likely it is for an image to appear in the real world. While we think the unnatural test cases generated by MetaIC are still useful, we argue that IC systems are mainly used for natural images (\textit{e.g.}, assistance for visually-impaired people) and whether an IC system can provide correct captions for natural images is more important.
\DIFdelbegin 

\DIFdelend 
 \DIFdelbegin 

\DIFdelend

\section{Approach and Implementation}

\begin{figure*}[ht]
		\centering
		\includegraphics[width=0.98\linewidth]{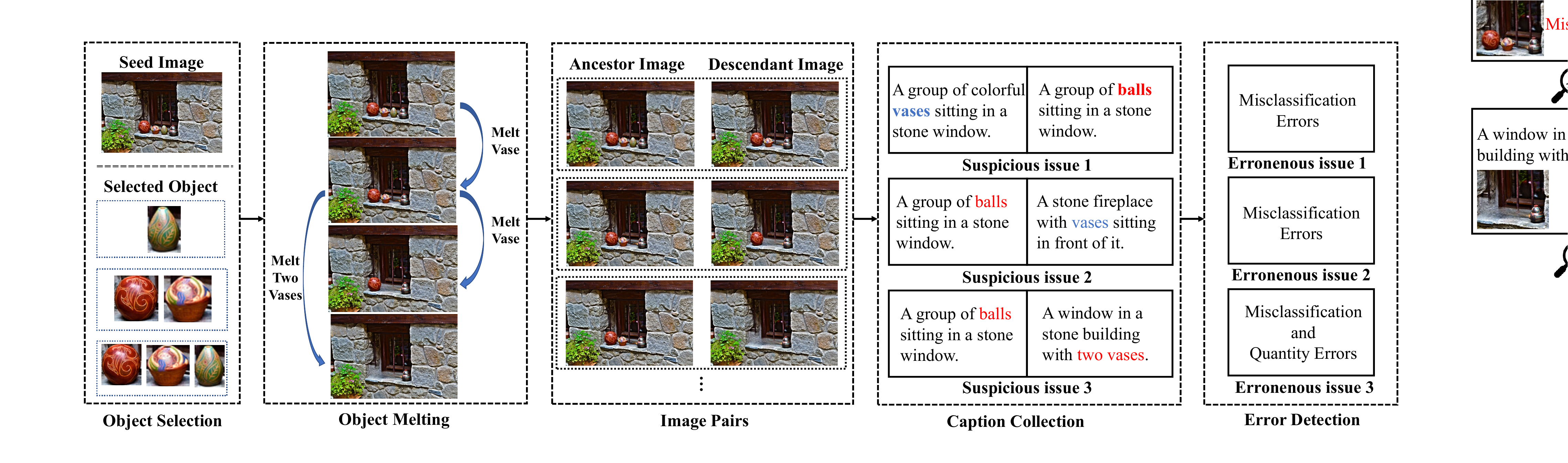}
		\caption{Overview of {\methodname}.}
		\label{fig:architecture}
\end{figure*}

This section introduces \textit{\underline{r}ecursive \underline{o}bject \underline{me}lting} ({\methodname}) and its implementation details. 
Our key insight is that the object set of an image after melting should be the subset of that of the original image. 
The input of {\methodname} is a list of images and the output is a list of suspicious issues, where each issue contains an \textit{ancestor}, a \textit{descendant} (image after melting object(s) in the ancestor), and their corresponding captions returned by the software under test. 
Fig.~\ref{fig:architecture} provides an overview of {\methodname}, which carried out the following steps:

\begin{itemize}

\item \textit{Object selection}. We recursively select the objects in the original image to be removed, which constructs the image pairs of ancestors and descendants.

\item \textit{Object melting}. We melt the selected objects by removing the objects and inpainting the missing part.

\item \textit{Caption collection}. We feed the image pairs to the IC system under test and collect their captions.

\item \textit{Error detection}. We focus on object names in the captions of ancestors and descendants. If any metamorphic relation is violated, {\methodname} will report a suspicious issue.
\end{itemize}

\begin{algorithm}[t]
	\small
	\caption{ An implementation of object selection }
	\label{alg:Rome}
	\begin{flushleft}
		\textbf{Input:} a maximum recursion depth $max\_depth$, a list of available object candidates $objects$ in a seed image \\
		\textbf{Output:} $list\_selected$, a list containing  pairs of object set $(ances\_{set}, desc\_{set})$, where $ances\_{set}$ and $desc\_{set}$ represent the objects selected to be removed in the ancestor image and descendant image, respectively
	\end{flushleft}
	\begin{algorithmic}[1]
		\State $selection\_list\leftarrow List(\hspace{0.5ex})$
		\Comment{Initialize with empty list}
		\State $ances\_set\leftarrow Set(\hspace{0.5ex})$
		\Comment{Initialize with empty set}
		\State $desc\_dict\leftarrow Dict(\hspace{0.5ex})$
		\Comment{Initialize with empty Dict}
		
		\State \Call{RECURSION}{$ances\_set$}
		\Statex 
        \Function{RECURSION}{$ances\_set$}
        \State $ances\_index\leftarrow Hash(ances\_set)$
		\Comment{Step 1}
		\If{$ances\_index$ in $desc\_dict.keys()$}
		\State \Return
		\EndIf
		\State $desc\_dict[ances\_index]\leftarrow List()\hspace{0.5ex}$
		\State $desc\_dict[ances\_index].append(ances\_set)\hspace{0.5ex}$
        \If{ $ances\_set.size$ $\geq$ $max\_depth$ }
		\State \Return
		\EndIf
		\ForAll{$obj$ in $(objects \setminus ances\_set)$}
		\Comment{Step 2}
		\State $next\_set\leftarrow Copy(ances\_set)\hspace{0.5ex}$
		\State $next\_set.add(obj) \hspace{0.5ex}$
		\State RECURSION($next\_set$)
		\State $next\_hash \leftarrow Hash(next\_state)$
        \ForAll{$desc\_set$ in $(desc\_dict[next\_hash])$}
        \If{ $desc\_set$ not in $desc\_dict[ances\_index]$ }
		\State $desc\_dict[ances\_index].add(desc\_set) \hspace{0.5ex}$
		\EndIf
		\State $object\_pair\leftarrow (ances\_set, desc\_set)\hspace{0.5ex}$
		\State $selection\_list.append(object\_pair) \hspace{0.5ex}$
        \EndFor
        \EndFor
        \State \Return $selection\_list$
        \EndFunction

		
	\end{algorithmic}
\end{algorithm}

\subsection{Object Selection}

\begin{figure}[ht]
		\centering
		\includegraphics[width=0.72\linewidth]{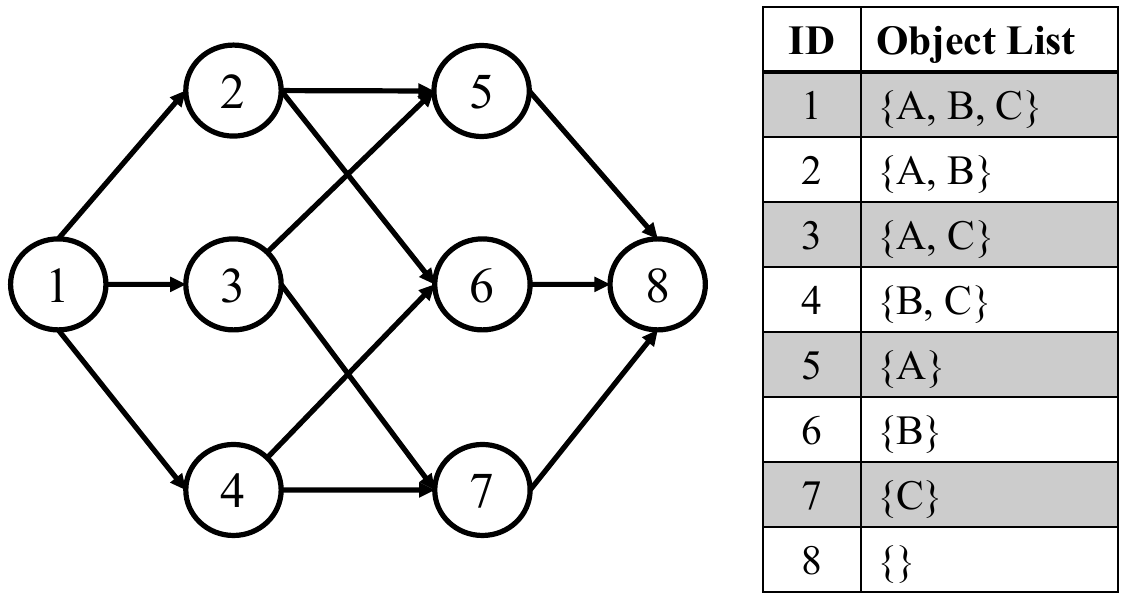}		\caption{An example of recursive object selection. The original image in this example contains three objects: A, B, and C.}
		\label{fig:dag}
\end{figure}

The goal of this step is to construct pairs of images, where a descendant is melted from its corresponding ancestor.
For images with multiple objects, we recursively select the objects to be removed.
Thus, a generated image could be the ancestor of another image with more objects removed.
For example, in Fig.~\ref{fig:rome_recur_show}, the original image is the ancestor of the other three images, and the second image is also the ancestor of the third image.
To demonstrate the recursive selection process, we present a directed acyclic graph (DAG) in Fig.~\ref{fig:dag} as an example.
Suppose there are three objects in the original image, we have $2^3=8$ states where each node denotes a state and each edge denotes the removal of a certain object.
Node 1 represents the seed image in which there are three objects: A, B, and C. If we use node 1 as the ancestor, it has $2^3-1$ descendants.
If we remove one object, \textit{i.e.}, A, B, or C, we arrive at nodes 4, 3, and 2, respectively, each of which has $2^2-1$ descendants.
Similarly, we could further remove one more object and get nodes 5, 6, and 7, where there is only one object left and each node only has one descendant.

To formalize the process of recursive object selection, we present its pseudocode in Alg.~\ref{alg:Rome}.

\textbf{Step 1}: We denote $ances\_set$ as the set of objects selected to be deleted in the current state and $desc\_dict$ as the dictionary to record the descendant states of the ancestor's state.
First, we assign a hash code for an $ances\_set$ (line 6) and it is used as the index in $desc\_dict$.
Then we initialize the descendant states of the current state (line 7-12) and return if the state has been visited (line 7-8), or the recursion reaches the maximum depth (line 11-12).
In our experiment, we created a large number of test cases (9,250 test pairs) using only 226 seed images, by setting the maximum recursion depth $max\_{depth}$ as 2.

\textbf{Step 2}: 
In step 2, we create a variable $next\_set$ and record the current state (line 15). Then we update the state by adding the selected object (line 16).
It calls the recursion function to add the descendant states and collects the pairs of object sets $(ances\_{set}, desc\_{set})$ (line 17-23).


In general, for an image with n objects, we can generate $2^n-1$ descendants and form using one seed image by recursively iterating the available objects.
One ancestor image $I_a$ and one descendant image $I_d$ construct an image pair $(I_a, I_d)$.
For an original image with n objects, if we use this image as the ancestor, we can generate $2^n-1$ image pairs.
Using the synthesized image which has removed $i$ object, we can generate $\binom{n}{i} \times (2^{n-i} - 1)$ image pairs.
By summing all the pairs, the total number of pairs is:
\begin{equation}
    N_{caps} = 3^n - 2^n,
\end{equation}
If one seed image has eight available objects, we can generate 6,305 image pairs.
Therefore, in theory, we can use the {\methodname} to generate a large number of test cases.

\subsection{Object Melting}
Next, based on the objection selection results, we use an image inpainting technique for object melting.
The goal of image inpainting is to realistically fill in some target areas in an image, which requires the inpainting model to understand the structural and textual information of natural images.
In {\methodname}, we adopt LaMa~\cite{suvorov2022resolution} as our image inpainting model, which adopts Fast Fourier convolutions (FFCS) to enhance its inpainting ability.

Specifically, we feed the image and the binary mask to LaMa, marking the object areas to be removed from the source image.
Then LaMa removes the objects and fills the missing areas by leveraging its neighboring structural and textual information.
In Fig.~\ref{fig:rome_recur_show}, we mask the vases in the image and LaMa successfully fills the missing areas with possible background contexts.
However, when the area to be filled becomes too large, the performance of LaMa is reduced.
To prevent this, we avoid selecting the largest object in the original image for melting.



\subsection{Caption Collection}
Now we have generated a large number of images, which can construct image pairs with their ancestors.
All the image pairs are fed into IC system under test to collect captions.
In this paper, we test one paid IC service, \textit{i.e.}, Microsoft Azure Cognitive Services~\cite{miocrosoft_azure_api}, and four representative IC algorithms, which are Attention~\cite{xu_show_2015}, Oscar~\cite{li_oscar_2020}, VinVL~\cite{zhang_vinvl_2021}, and OFA~\cite{wang2022unifying}.

\subsection{Error Detection}
After caption collection, {\methodname} inspects the metamorphic relations (MRs) of each caption pair $(C(I_{a}), C(I_{d}))$.
We design two MRs for {\methodname}: object melt and class melt.
If any of the metamorphic relations was violated, the pair and the corresponding captions will be reported as a suspicious issue.

\textbf{MR1 Object Melt.} We denote the IC system as $C$, the ancestor image as $I_{a}$, the descendant image as $I_{d}$, the deleted objects as $obj_{del}$.
The caption pair produced by $C$ are $C(I_{a})$ and $C(I_{d})$.
$S(C(I_{a}))$ denotes the set of object classes in $C(I_{a})$, $S(C(I_{d}))$ indicates the set of object classes in $C(I_{d})$. \textit{MR1} is defined as follows:
\begin{equation}
    \begin{aligned}
    S(C(I_{d})) \subseteq S(C(I_{a})), \\
    S(C(I_{a})) \setminus S(C(I_{d})) \subseteq S(obj_{del}),
    \end{aligned}
\end{equation}
which means that the object set of a descendant should be the subset of that of its ancestor, and the difference between their objects should only come from the deleted objects.



\textbf{MR2 Class Melt.} If there is only one object of a class, we will denote it as $obj\_alldel$, which means there is no such class in the image after removing the object.
MR2 is designed to handle the cases where we remove the single object in the ancestor image, which is defined as below:
\begin{equation}
    S(C(I_{d})) \subseteq S(C(I_{a})) \setminus S(obj_{alldel}).
\end{equation}




Our MRs mainly check the names in the caption. When checking object names that have synonyms and consist of multiple words, we cannot trivially conduct string comparison. To this end, we develop  heuristic rules as follows.

\textbf{Rules for word groups.}
Some object classes have names that are made up of multiple words, such as hot dog, bow tie, and bowl sink.
In this study, we focus on the 80 classes of COCO Caption~\cite{chen2015microsoft}, so we have rules for the two cases: (1) the entire word group is an object class in COCO Caption, (2) the entire word group is not object class in COCO Caption.

Rule 1: If the complete word group is an object class in COCO Caption, we add the whole word group to the set of object classes in a caption and exclude the single word from the word group.
For instance, "hot dog" is a class in COCO Caption. When iterating through the classes in a caption, we initially add "dog" and "hot dog" to the set of object classes. Then, we remove "dog" because it is part of the word group "hot dog," to eliminate ambiguity.


Rule 2: If the word group is not a keyword, we only keep the later noun in the set of the object classes.
For example, "bowl sink" and "bow tie" are two word groups not in COCO Caption.
The first word of each is a qualifier for the second word in them, i.e., "bowl" for "sink", and "bow" for "tie."
Therefore, we only add the "sink" and "tie" into the set of object classes.

\textbf{Rule for synonyms.}
COCO Caption contains 80 object classes, each has a keyword to denote the classes.
A word may have many synonyms, therefore different words can be used to describe the same object in the captions.

We denote the function to search for hypernyms as $Hyper()$, and the function to search for denotations as $Denot()$.
If the word $w$ is in the keyword of COCO Caption, we can use it directly to check metamorphic relations.
Otherwise, we leverage different denotations of the word and their hypernym to map it into keywords in COCO Caption.
The rule to map the word into the keyword of COCO Caption is implemented in Alg.~\ref{alg:syn_rule}, including two steps:

\textbf{Step 1.} We first generate a denotation list $denot\_list$ to record the denotations of the word.
If any word in $denot\_list$ is the keyword in $keyword\_cap$, we return that word (line 7-9).
Otherwise, we search for the hypernyms of each word in $denot\_list$ (line 10-16).
In our experiment, we set the maximum depth $D$ as 3 and it performs well in testing the five IC systems. 

\textbf{Step 2.} In step 2, we traverse the word in $denot\_list$ and add the hypernym of the word to $hyper\_list$ (line 22).
Meanwhile, we call function $Denot(word)$ to get the denotations of these words (line 23).
If any word of $Denot(word)$ is in the caption pair $(C(I_{a}), C(I_{d}))$, we return the word and $hyper\_list$ (line 23-26).

With its rules for word groups and synonyms, {\methodname} has the advantage of being able to test images with object classes that have names longer than two words.
However, the SOTA IC testing algorithm, MetaIC is only capable of testing images with object classes that have single-word names, limited to only 60 such classes.





\begin{algorithm}[h]
	\small
	\caption{ An implementation of rules for synonyms}
	\label{alg:syn_rule}
	\begin{flushleft}
		\textbf{Input:} a word $input\_word$, a keyword set $keyword\_cap$ in caption pair $(C(I_{a}), C(I_{d}))$, a keyword set of COCO Caption KEYWORD, a maximum iteration depth $D$ \\
		\textbf{Output:} $mapped\_word$, which the word $input\_word$ is mapped to
	\end{flushleft}
	\begin{algorithmic}[1]
        \Function{Mapping}{$input\_word$}
        \If{$word$ is in KEYWORD}
        \State \Return $word$
        \EndIf
        \State $cur\_iter \leftarrow 0$
        \Comment{Step 1}
        \State $denot\_list \leftarrow Denot(word)$
        \State $find\_word \leftarrow None$
        
        \ForAll{$ele$ in $denot\_list$}
        \If{$ele$ is in $keyword\_cap$}
        \State \Return $ele$
        \EndIf
        \EndFor
        
        \While{$cur\_iter < D$}
        \State $cur\_iter \leftarrow cur\_iter + 1$
        \State $find\_word, hyper\_list \leftarrow \Call{Synonyms}{denot\_list}$
        \If{$find\_word$ is not $None$}
        \State \Return $find\_word$
        \Else
        \State $denot\_list \leftarrow hyper\_list$
        \EndIf
        \EndWhile
        \State \Return $find\_word$
        \EndFunction
        
        \Function{Synonyms}{$denot\_list$}
        \State $find\_word \leftarrow None$
        \Comment{Step 2}
        \State $hyper\_list \leftarrow List()$
        \ForAll{$word$ in $denota\_list$}
        \State $hyper\_list.append(Hyper(word))$
        \ForAll{$ele$ in $Denot(word)$}
        \If{$ele$ is in $keyword\_cap$}
        \State \Return $ele, hyper\_list$
        \EndIf
        \EndFor
        \EndFor
        \Return $find\_word$, $hyper\_list$
        \EndFunction
        
	\end{algorithmic}
\end{algorithm}

\section{Evaluation}
In this section, we evaluate {\methodname} on five IC systems, including 4 representative IC models and one commercial API (i.e., Microsoft Azure Cognitive Service). 
Specifically, we aim to answer the following research questions (RQ):
\begin{itemize}
   \item RQ1: How natural is the image in the test case generated by {\methodname}?
   \item RQ2: How effective is {\methodname} at finding erroneous issues? 
   \item RQ3: What kinds of captioning errors can our approach find?
   \item RQ4: How can we utilize {\methodname} to improve the IC systems?
\end{itemize}

\subsection{Experimental Setup}

\subsubsection{Experimental Environments and Implementation}
All experiments are conducted on a Linux (Ubuntu 20.04.2 LTS) workstation with  64GB Memory and GeForce RTX 3090 GPU.
For POS tagging, we use the XPOS implemented in Stanza~\cite{stanza}, an NLP package powered by Stanford NLP Group.
To search for a word's denotation and hypernym, we use Wordnet, a lexical database provided by Princeton University and implemented in NLTK~\cite{nltk}.
All the code and datasets we used will be open-sourced.

\subsubsection{Comparative Methods}
We compare our proposed {\methodname} to MetaIC~\cite{yu2022automated}, the current SOTA framework for testing image captioning systems. MetaIC operates as follows: (1) It extracts objects from images and constructs a standalone object pool.
(2) From the pool, an image is randomly chosen and inserted into background images to produce synthesized images with varying degrees of overlap.
(3) An image pair consisting of a background image and a synthesized image is created, and the pair is fed into the IC system to collect a caption pair.
(4) MetaIC reports any suspicious issues using metamorphic relations. MetaIC can run in two different modes: \textit{non-overlapping} (MetaIC-NO) and \textit{partial-overlap} (MetaIC-PO), where the overlap ratio can be 15\%, 30\%, or 45\%.
For a fair comparison, we use both non-overlapping and partial-overlap modes. We choose the 30\% overlap ratio as the default for the partial-overlapping mode due to its superior performance.


\subsubsection{Dataset}
We evaluate {\methodname} on MS COCO Caption~\cite{chen2015microsoft}, a standard and most-popular dataset in image captioning, which contains 123,000 images.
Specifically, we follow the publicly available splits\footnote{https://cs.stanford.edu/people/karpathy/deepimagesent/} to use 5,000 images for both validation and testing.
For the baseline method MetaIC, we follow the work~\cite{yu2022automated} to leverage an additional dataset Flickr~\cite{flickr} as the object image.


\subsubsection{Benchmark IC Systems}
Our evaluation is based on five benchmark IC systems, including four representative IC models (i.e., Attention~\cite{xu_show_2015}, Oscar~\cite{li_oscar_2020}, VinVL~\cite{zhang_vinvl_2021}, OFA~\cite{wang2022unifying}) and Microsoft Azure Cognitive Service API~\cite{miocrosoft_azure_api} (in short, MS Azure API). 
For the Oscar model, we obtained it by fine-tuning on the COCO dataset and reproduced the results presented in the paper~\cite{li_oscar_2020}. 
For the remaining ones, we directly used the well-trained models released by the authors. 
Except for the Attention model, all other models we used are the base model.


\subsection{Naturalness Evaluation on Images Synthesized by {\methodname} (RQ1)}
While IC has many applications, we believe captioning natural images is the most important because many critical real-world scenarios rely on it, such as guiding the visually-impaired people~\cite{assistVisualImpaired} or helping the explorer analyze the landscape~\cite{arcgis_api}.
In this part, we target to evaluate how natural the images synthesized by {\methodname} are.
As introduced in Section~\ref{subsec:motivating_example}, the "naturalness" of the generated images is critical to the testing of IC systems, which refers to the likelihood of the images appearing in the real world.
To this end, we conduct a human evaluation study on four groups of images and compare the results.
The four groups of images are: raw images in the MS COCO Caption~\cite{chen2015microsoft} dataset, and synthesized images from {\methodname}, MetaIC-NO and MetaIC-PO~\cite{yu2022automated}.
We select 1,000 images from each group for the evaluation.
The images synthesized by our {\methodname} and baseline methods are based on the same set of 226 seed images.

The evaluation standard is defined as follows: the \textit{naturalness score} ranges from "1" (very unnatural) to "4" (very natural) and a higher score means that the image is more natural and thereby better. Specifically, the meaning of each score is shown as below:
\begin{itemize}
    \item “4” denotes that the image appears to have been captured in a natural setting and appears to be a true-to-life representation of nature.
    \item “3” denotes that the image may not be entirely natural, but it could still have been captured in nature.
    \item “2” denotes that the image is somewhat unnatural and would be difficult to capture in nature.
    \item “1” denotes that the image appears to be highly unnatural and cannot be considered a representation of nature.
\end{itemize}



\paragraph{Crowd-sourcing}
\begin{figure}[ht]
		\centering
		\includegraphics[width=0.9\linewidth]{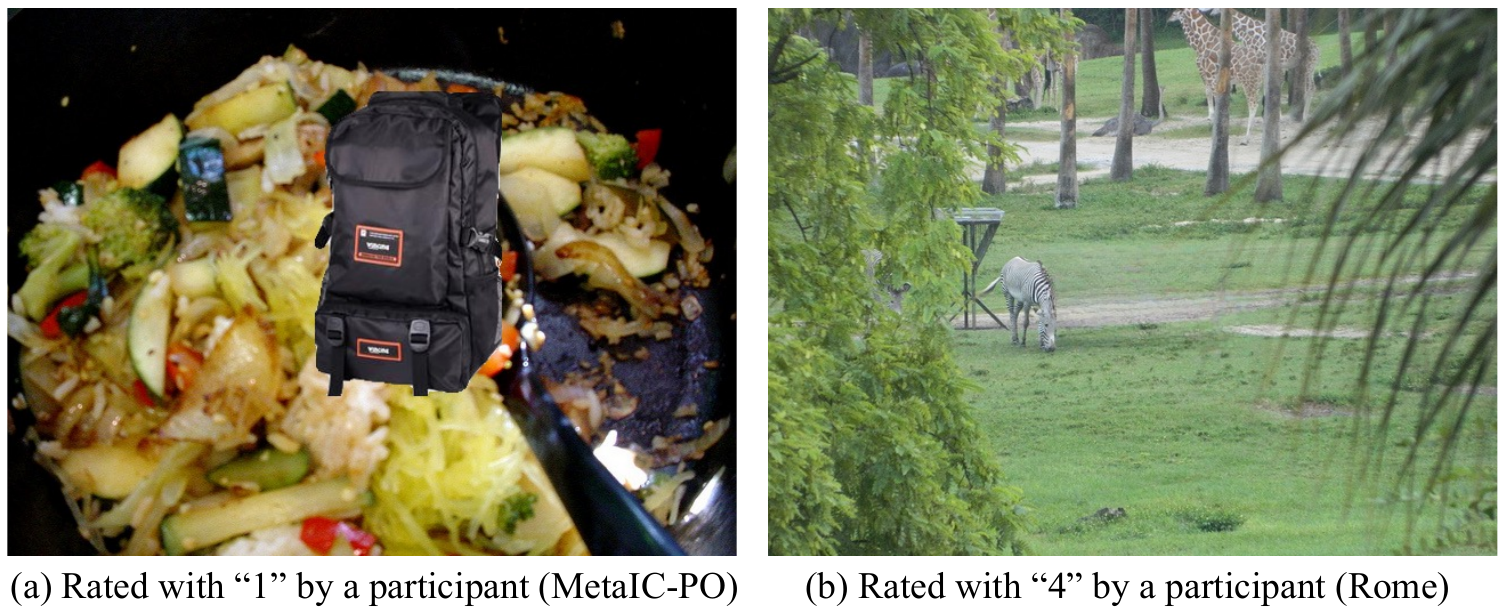}
		\caption{Examples rated by the participants in the user study.}
		\label{fig:rater}
\end{figure}
We recruited 24 people to rate the \textit{naturalness score} for the images on Prolific~\cite{prolific}, a platform to post tasks and hire workers.
We divide 4,000 images into 8 sessions, and assign three participants to each session.
To ensure the inter-rater reliability of our user study, we calculate the Intraclass Correlation Coefficients (ICC)~\cite{koo2016guideline} score to evaluate the absolute agreement among multiple raters during the human evaluation. 
The ICC score of the human evaluation is 0.83, which indicates excellent reliability of the user study.

Fig.~\ref{fig:rater} presents two images in the user study, where the food with a tiny bag on its top (Fig.~\ref{fig:rater}-(a)) generated by MetaIC-PO is rated by a participant with "1", and 
the natural image of animals in a field (Fig.~\ref{fig:rater}-(b)) generated by {\methodname} is rated by a participant with "4".
The naturalness scores for MS COCO Caption Dataset, {\methodname}, MetaIC-NO, and MetaIC-PO are 3.56, 3.33, 1.41, and 1.37.
We can observe that MS COCO Caption Dataset gets the highest naturalness score (3.56), which is intuitive because images in this group are real-world images.
The images synthesized by {\methodname} achieves a high score of 3.33, which is only 0.23 score lower than the raw images.
The high score indicates that the recruited raters think most of our synthesized images are close to real-world images, demonstrating the strong ability of {\methodname} in generating natural test cases.
On the contrary, the naturalness scores of images synthesized by two MetaIC methods are much lower.
In detail, MetaIC-NO and MetaIC-PO achieve naturalness scores of only 1.41 and 1.37, respectively.
It shows that the raters believe most images generated by MetaIC are unnatural.

\begin{center}
\fbox{
\parbox{0.9\linewidth}{\textbf{RQ1 Answer}: Images generated by {\methodname} are much more natural than the ones generated by MetaIC.
The naturalness score of {\methodname} shows that its synthesized test cases are very close to natural images.}
}
\end{center}

\begin{table*}[h]
    \centering
    \caption{Precision of {\methodname} and baseline methods}
    \resizebox{0.70\linewidth}{!}{
    \begin{tabular}{c|c|c|c|c|c}
    \hline & \multicolumn{5}{|c}{ IC Testing Approaches  } \\
    IC systems & {\methodname} & {\methodname} (MR1) & {\methodname} (MR2) & MetaIC-NO & MetaIC-PO  \\
    \hline  
    OFA~\cite{wang2022unifying} &{91.08 (1634/1794)}  & \textbf{96.64} (632/654)  & 88.84 (1130/1272)  & 89.07 (725/814) & 88.74 (717/808)\\
    
    Oscar~\cite{li_oscar_2020} &92.17 (1824/1979) & \textbf{97.32} (907/932)  & 89.20 (1123/1259) & 91.45 (749/819) & 90.01 (739/821) \\
    VinVL~\cite{zhang_vinvl_2021} &88.47 (1673/1891) & \textbf{93.70} (862/920)  & 85.38 (987/1156) & 87.80 (655/746)  & 87.32 (654/749) \\
    Attention~\cite{xu_show_2015} &86.47 (2320/2683) & 97.14 (1360/1400)  & 78.42 (1214/1548)  & \textbf{98.98} (967/977) & 98.87 (961/972) \\
    
    MS Azure API~\cite{miocrosoft_azure_api} & 88.13 (1670/1895) & 93.33 (951/1019)   & 84.67 (895/1057) & \textbf{97.68} (928/950) & 97.56 (920/943) \\
    \hline
    \end{tabular}
    \label{tab:precision}
    }
\end{table*}

\subsection{Effectiveness of Finding Erroneous Issues (RQ2)}
In this section, we evaluate the effectiveness of {\methodname} in identifying erroneous issues for IC systems. Automated testing approaches like {\methodname} can  generate a pair of images (i.e., $I_{a}, I_{d}$) that may trigger the IC system errors.
Toward the goal of evaluating effectiveness, manual labels based on the outputs of these approaches are needed. 
Specifically, we denote a suspicious issue as a pair of captions generated by IC systems, that is, $p=(C(I_{a}), C(I_{d}))$. 
If any caption in the pair $p$ is confirmed to have errors by humans, we then say $p$ is an erroneous issue (denoted as $p_{error}$). Given an IC system, automated testing approaches such as {\methodname} can produce a list of suspicious issues, among which some are confirmed to be erroneous issues. Therefore, the effectiveness of an automated testing approach can then be measured by:

\begin{equation}
    Precision = \frac{|\{p_{error}\}|}{|P|},
\end{equation}
where $P$ is the set of all suspicious issues reported by an automated testing approach and $\{p_{error}\}$ is the set of erroneous issues confirmed by humans.  $|X|$ is the size of a set $X$.

To confirm whether a suspicious issue is erroneous, two co-authors check if the generated captions contain some errors and then categorize them.
The user-evaluation results have a Cohen's kappa of 0.866, showing a substantial level of agreement~\cite{mchugh2012interrater}.
Note that the manual labels here are solely used to measure whether a testing approach can precisely find erroneous issues, while the method itself is fully automated and does not require human intervention. A low precision indicates that  although the testing approach can generate many suspicious issues, only a few of them can make the IC systems fail. 


    
    

We evaluate the precision of  {\methodname} as well as two baseline methods MetaIC-NO and MetaIC-PO based on the aforementioned five benchmark IC systems.
As shown in Table~\ref{tab:precision}, {\methodname} performs well on all five IC systems with a precision score ranging from 86.47\% to 92.17\%. 
Moreover, compared with MetaIC-NO and MetaIC-PO, {\methodname} achieves higher precision on three IC systems including OFA, Oscar, and VinVL.
However, {\methodname} achieves lower precision on Attention and MS Azure API than the comparative methods. 
We find that Attention and MS Azure API cannot generate comprehensive captions if the input image is not natural. MetaIC inserts an object on the background image unnaturally, where description about the object is often ignored by the two IC systems. Therefore, MetaIC tends to identify the case as an error, leading to its high precision score. Considering that the images generated by {\methodname} are much more natural than MetaIC, the uncovered captioning errors of {\methodname} yield superior reference value for the captions of real-world images.


    


In {\methodname}, there are two metamorphic relations (MRs).
To show the effectiveness of the two MRs in finding erroneous issues, we conduct an ablation study.
That is, we compare the {\methodname} with only MR1 encoded, MR2 encoded and {\methodname} with both MR1 and MR2 (denoted as MR1+MR2).
As shown in Table~\ref{tab:precision}, MR1 and MR2 can both achieve high precision, where MR1 has higher precision.
By adopting both of them, we can report more erroneous issues than by using single MR1 or MR2 while keeping the high precision (86.47\%-92.17\%).



\begin{center}
\fbox{
\parbox{0.9\linewidth}{\textbf{RQ2 Answer}: {\methodname} has achieved a high precision (86.47\%-92.17\%) for testing the five IC systems.}
}
\end{center}


\begin{figure*}[ht]
		\centering
		\includegraphics[width=0.94\linewidth]{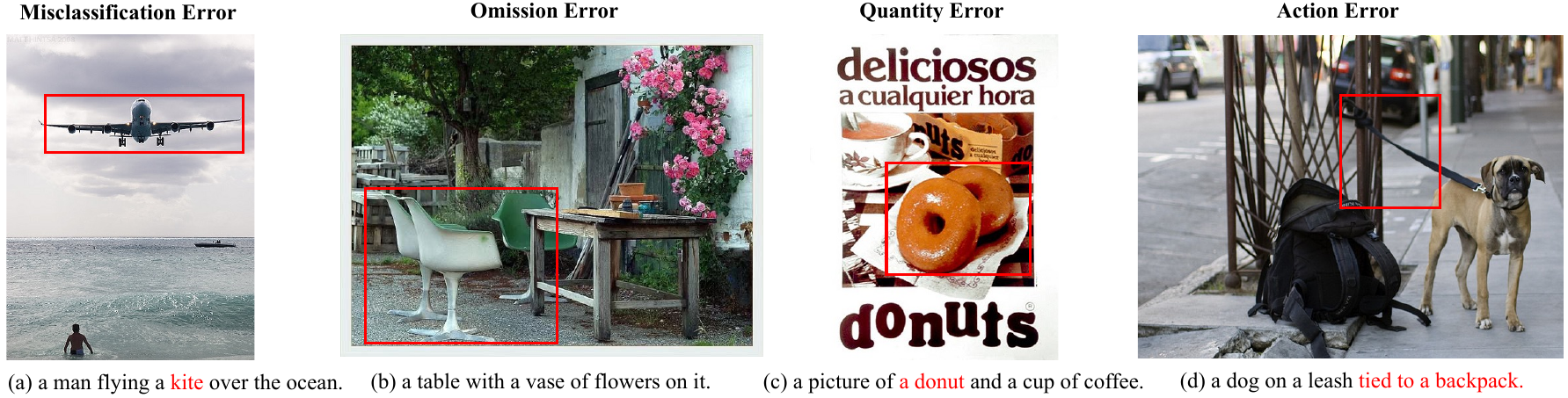}
		\caption{Examples of captioning errors reported by {\methodname}.}
		\label{fig:tp}
\end{figure*}

\subsection{Categories of Reported Captioning Errors (RQ3)}
In this section, we conduct analysis on the detected captioning errors to understand the error behaviors. To this end, we categorize the erroneous issues reported by {\methodname} into different error types. For easy understanding, Fig.~\ref{fig:tp} shows samples for each error type.
In general, there are four categories of errors with details introduced below. 
\begin{itemize}
    \item \textbf{Misclassification error} is an error that IC systems mis-classify an object in the image as another object. Fig.~\ref{fig:tp}-(a) demonstrate a misclassification errors of mistakenly recognizing an airplane as a kite.
    \item \textbf{Omission error} is an error that IC systems omits some objects of the images. Fig.~\ref{fig:tp}-(b) presents an omission error where the chairs in the image has been ignored.
     \item \textbf{Quantity error} is an error that IC systems count the wrong quantity of objects in the caption. Fig.~\ref{fig:tp}-(c) shows a quantity error where two donuts are mis-counted as one.
    \item \textbf{Action error} is an error that IC systems mistakenly describe the relationship of the objects. Fig.~\ref{fig:tp}-(d) presents an action error that the dog is mistakenly described to be tied to a backpack.
\end{itemize}

We have found 9,417 captioning errors from the erroneous issues reported by {\methodname}, where the misclassification, omission, quantity, and action errors comprise 41\%, 34\%, 23\%, and 2\% of the total, respectively.

\begin{center}
\fbox{
\parbox{0.9\linewidth}{\textbf{RQ3 Answer}: {\methodname} can find four categories of captioning errors, including misclassification error, quantity error, omission error and action error, which comprise 41\%, 34\%, 23\%, and 2\% of the total, respectively.}
}
\end{center}

\subsection{Usage Scenarios of {\methodname}  (RQ4)}
In early sections, we demonstrated that {\methodname} can automatically identify captioning errors with high precision for various IC systems.
This raises a question: apart from error identification, what other benefits does {\methodname} offer to IC systems?
To address this question, we identified two scenarios where {\methodname} could prove advantageous. 
The first scenario is to debug the training data while the second scenario is to correct (or improve) the current IC systems.

\subsubsection{Finding Labeling Errors of Training Corpus}
The outstanding of AI and machine learning highly relies on the dataset quality, including COCO Caption~\cite{chen2015microsoft}, ImageNet~\cite{krizhevsky2012imagenet}, MNIST~\cite{deng2012mnist}, etc.
A recent study~\cite{mit_findErrors} has identified significant labeling errors in popular AI benchmark datasets, highlighting the possibility of errors in these datasets.
To uncover labeling errors in COCO Caption, we used {\methodname} to verify whether the objects in the images were correctly captioned in the ground truth labels.

During metamorphism, the two captions $C(I_a)$ and $C(I_d)$ may have common object classes, which are essential image features and are referred to as invariant object classes.
To verify the ground truth labels, we utilize the set of invariant object classes from the caption pair generated by OFA~\cite{wang2022unifying}.
Our rationale is that human workers are prone to writing typos or losing internet connectivity while labeling, but machine algorithms are less likely to make such errors.
If the set of invariant object classes $Set_{invar} = S(C(I_{a})) \cap S(C(I_{d}))$ is not present in the ground truth $GT$ of $I_{a}$, there might be labeling errors in $GT$.
To verify this, we check if $Set_{invar}$ is a subset of $S(GT)$.
If this condition is not met, the ground truth label may contain errors.
For this experiment, we only use the images in COCO Caption's training corpus as $I_{a}$.

\begin{figure}[ht]
		\centering
		\includegraphics[width=0.98\linewidth]{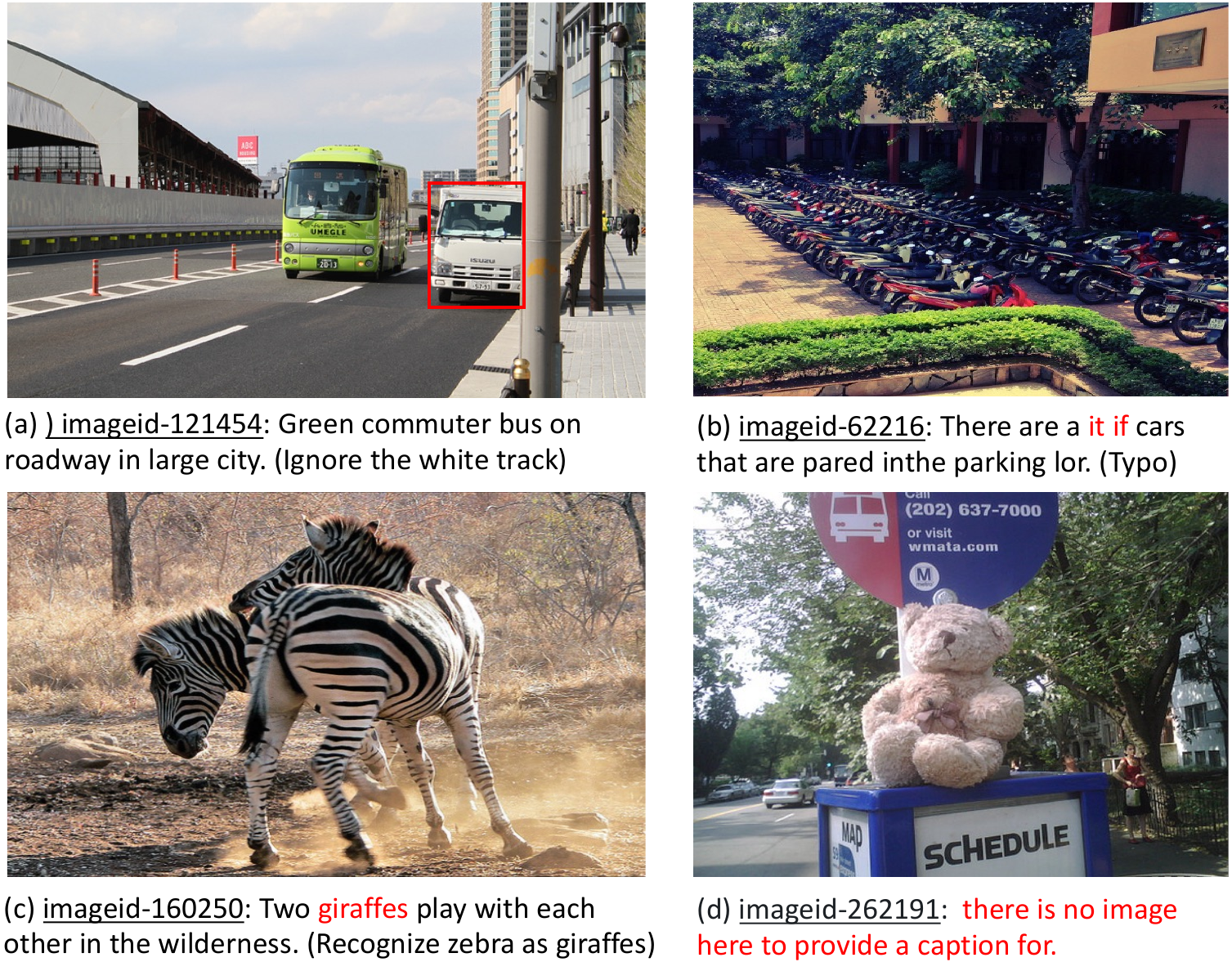}
		\caption{Labeling errors in COCO Caption~\cite{chen2015microsoft} training dataset.}
		\label{fig:gt_bugs}
\end{figure}

By using {\methodname}, we have found 219 labeling errors in COCO Caption, including four categories: (1) omission error, (2) typo, (3) misclassification error, (4) "no image" error.
The full list of "imageid" of these labeling errors is in the supplement material.
We can view the labelings of the labeling by searching the corresponding "imageid" in the website of COCO Dataset, \footnote{https://cocodataset.org}.
In Fig.~\ref{fig:gt_bugs}, we show representative examples of the labeling errors.
The errors like typos in Fig.~\ref{fig:gt_bugs}-(b) or labeling zebra as giraffes in Fig.~\ref{fig:gt_bugs}-(c) will severely impair the model training, bring along bias or other unstable factors.
Fig.~\ref{fig:gt_bugs}-(d) may be due to Internet errors which make the labeling workers unable to view the images.
Cleaning the IC dataset will be a promising research topic to improve the IC systems.

\subsubsection{Fine-Tuning with Errors Reported by {\methodname}}

To investigate whether we can utilize the erroneous captions reported by {\methodname} to improve the performance of IC systems, we re-label the erroneous captions found by {\methodname} following the labeling standard of COCO Caption~\cite{chen2015microsoft} and fine-tune Oscar~\cite{li_oscar_2020}.
In practice, we re-label 1,000 synthesized images generated from the training data of COCO Caption and use our re-labeled synthetic data as the augmented data in the training corpus.
By fine-tuning for 65K steps, we have achieved significant progress on the four evaluation metrics.
As shown in Table~\ref{tab:finetune}, BLEU-4 is increased by 1.0\%, METEOR is increased by 3.4\%, CIDEr is increased by 2.3\%, and SPICE is increased by 8.3\%.
According to a survey of IC systems (Table 2 in \cite{stefanini2021show}), our improvements on the four scores are significant.

\begin{table}[h]
    \centering
    \caption{Fine-tuning performance of {\methodname} on relabeled images }
    \resizebox{0.80\linewidth}{!}{
    \begin{tabular}{c|c|c|c|c}
    \hline & \multicolumn{4}{|c}{ Evaluation metrics  } \\
    IC systems & BLEU-4 & METEOR & CIDEr & SPICE  \\
    \hline
    Oscar & 40.5 & 29.7 & 137.6 & 22.8 \\

    Oscar$_{B-finetune}$ &\textbf{40.9} & \textbf{30.7} & \textbf{140.8} & \textbf{24.7} \\
    \hline
    Improvement (\%) & 1.0 & 3.4 & 2.3 & 8.3 \\
    \hline
    \end{tabular}
    \label{tab:finetune}
    }
\end{table}

\begin{center}
\fbox{
\parbox{0.9\linewidth}{\textbf{RQ4 Answer}: We can utilize {\methodname} to build robust IC systems by finding labeling errors in training data or improving  the model with the reported errorneus captions.
}}
\end{center}


\section{Studies on Real-world Applications}

\begin{figure*}[ht]
		\centering
		\includegraphics[width=0.97\linewidth]{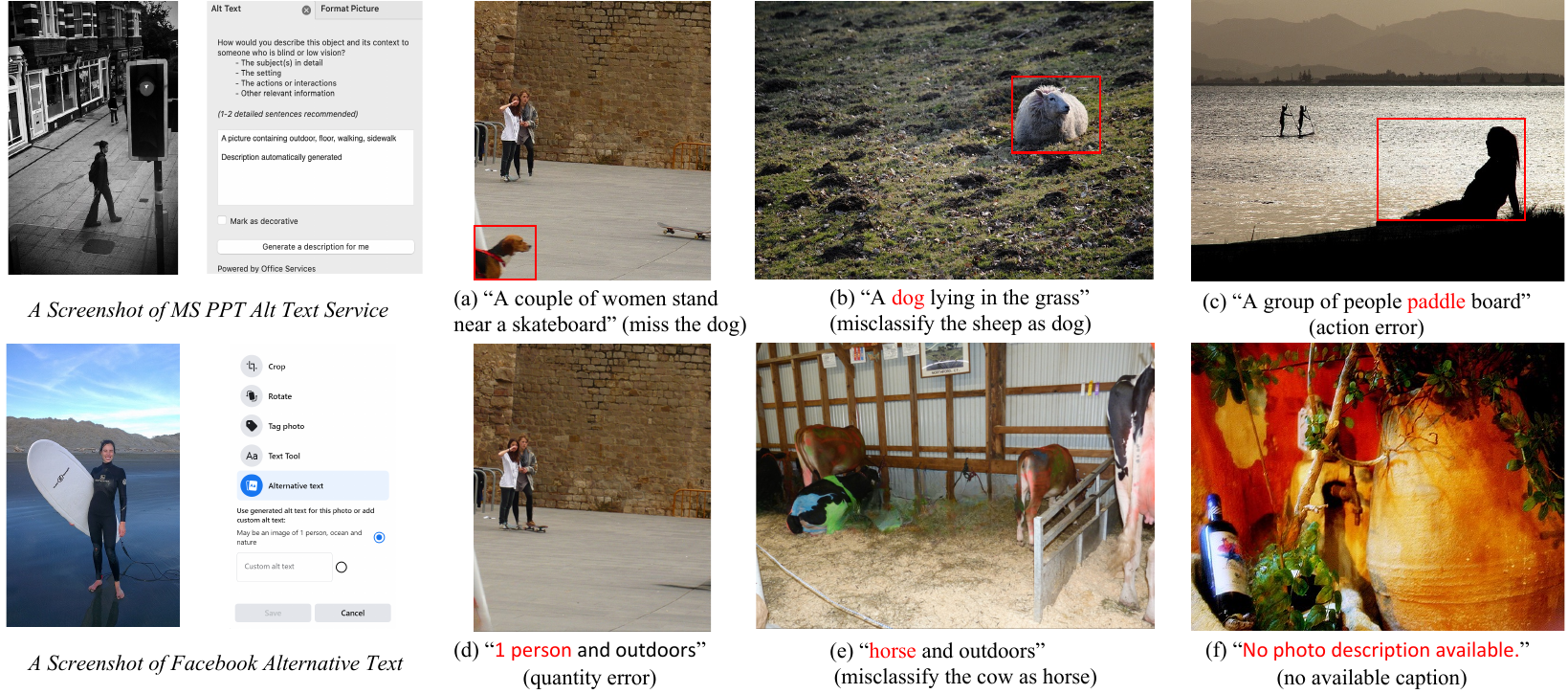}
		\caption{Captioning errors triggered by the Alternative Text Service in Microsoft Powerpoint (first row) and Facebook (second row).}
		\label{fig:fbms_bugs}
\end{figure*}


We have adopted {\methodname} to a commercial API, i.e., MS Azure API. 
While there are some other commercial services that integrate IC as an internal function, they do not provide  API interfaces.
To explore whether the captioning errors reported by {\methodname} also exist in these real-world applications, we randomly sample 100 images that trigger captioning errors of the four IC models and MS Azure API, and use them to manually test the performance of IC systems integrated in two popular commercial software applications: alternative text function in Microsoft Powerpoint~\cite{office365_api} and Facebook~\cite{facebook_api}.
Below, we present some examples of typical captioning errors produced by these two software applications.

\subsection{AI Captioning Errors in Microsoft Powerpoint}

Microsoft 365 has integrated automated IC systems in Microsoft Powerpoint, as shown in the leftmost image in Fig.~\ref{fig:fbms_bugs}, the first row.
To use this function, we only need to click the button "Generate a description for me".
In the first row of Fig.~\ref{fig:fbms_bugs}, we list some examples of erroneous captions found in Microsoft Powerpoint.
Fig.~\ref{fig:fbms_bugs}-(a) misses the dog in the image, while Fig.~\ref{fig:fbms_bugs}-(b) misclassifies the sheep as a dog.
Fig.~\ref{fig:fbms_bugs}-(c) assign the action "\textit{paddle}" to the woman that lies on the shore.
In our study on Microsoft Powerpoint, 46 images out of 100 images trigger captioning errors, where there are 29 omission errors, 13 misclassification errors, 3 quantity errors, and 1 action error.

\subsection{AI Captioning Errors in Facebook}
Facebook has adopted automated image captioning to enhance accessibility, which can benefit people who are blind or have low-vision.
As shown in the leftmost image in the second row of Fig.~\ref{fig:fbms_bugs}, Facebook's Automated Alt Text (AAT) technology can automatically create a description of the image.
In practice, it provides caption if the person who uploaded the image does not include alt text.
The caption generated by Facebook is combined with a prefix sentence "maybe an image of", For simplicity, we do not display the prefix.
Besides assisting people with visual diseases, alternative texts also serve as useful information for optimizing search engine optimization.
Therefore, automated alternative texts' ability to convey the information of the image is essentially important.
Otherwise, the search engine may recommend incorrect images to the users.

In the first row of Fig.~\ref{fig:fbms_bugs}, we list some examples of erroneous captions found in Microsoft Powerpoint.
In the second row of Fig.~\ref{fig:fbms_bugs}-(d) only describe one person in the image, while there are two people, and (e) misclassifies the cow as the horse.
Facebook AAT may fail to provide any caption, there is an example in In Fig.~\ref{fig:fbms_bugs}-(f).
In our study on Facebook AAT, 57 images out of 100 images trigger captioning errors of the alternative text function on Facebook, where there are 37 omission errors, 2 misclassification errors, 11 quantity errors, 1 action error, and 6 cases of no caption.
Our practice shows {\methodname}'s ability to unveil real-world software defects.



\section{Discussions}


\subsection{False Positives}
\begin{figure}[ht]
		\centering
		\includegraphics[width=0.90\linewidth]{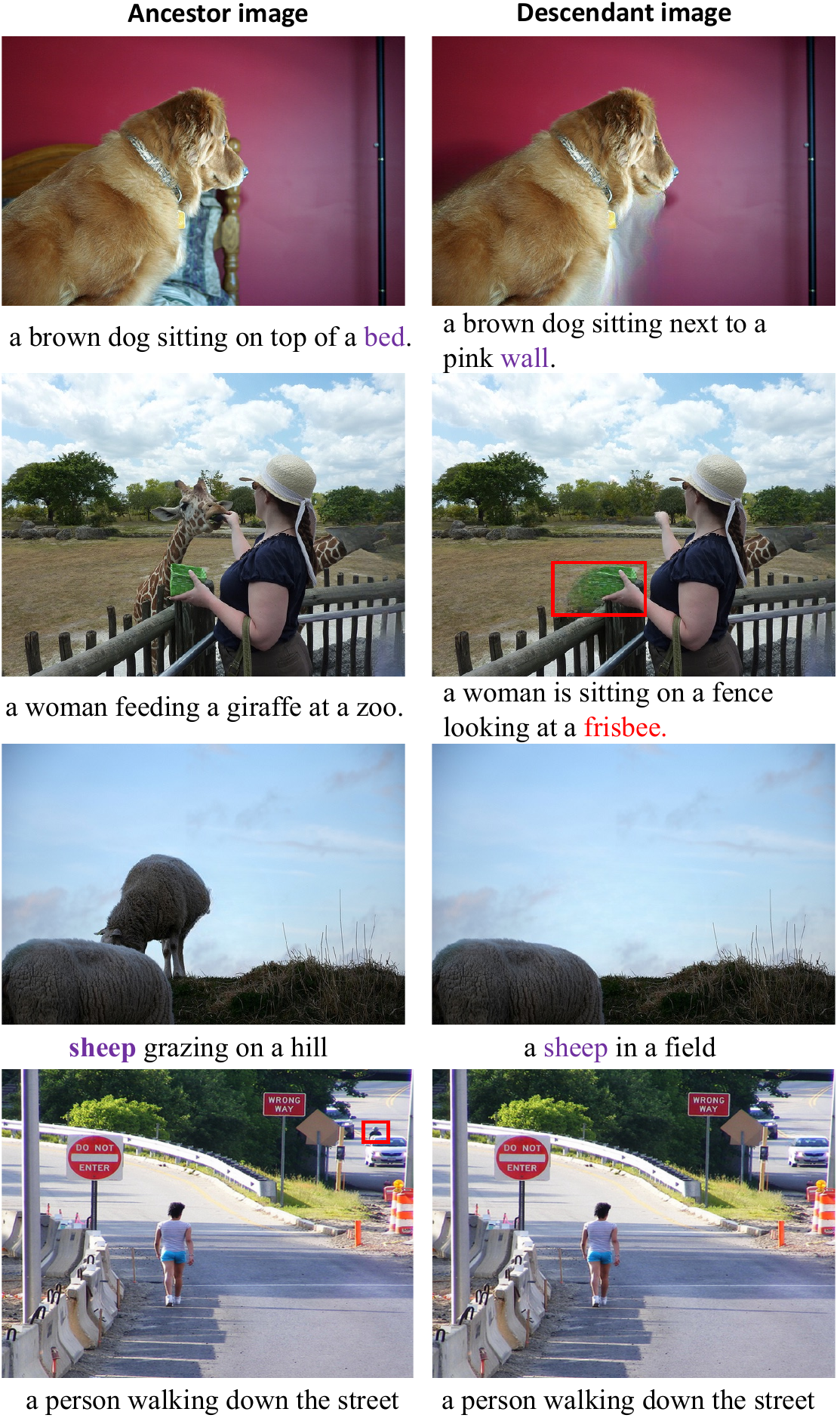}
		\caption{False positives produced by {\methodname}. }
		\label{fig:fp}
\end{figure}

Although {\methodname} has achieved a high precision, it mainly has four categories of false positives.
We show one example of each category in Fig~\ref{fig:fp}.
The false positive in the first row of Fig~\ref{fig:fp} is caused by the rule for synonyms of {\methodname}, which mistakenly assigns "bed" as the synonym of "wall".
Though {\methodname} can handle most of the synonyms, it can not handle all the cases.
We will adopt more rules and more advanced NLP tools to handle the synonyms in the future.
In the second row of Fig.~\ref{fig:fp}, the false positive is caused by the unsatisfactory result of image inpainting, which makes the basket in the person's hand hard to recognize.
In the third row of Fig.~\ref{fig:fp}, Stanza~\cite{stanza} mistakenly assigns the singular word form for the sheep in the left image, which gives birth to the false positive.
In the fourth row of Fig.~\ref{fig:fp}, {\methodname} remove the tiny person that is not significant and does not need to be described in the caption, which gives birth to the false positive.
There is only a rate of 10.9\% (1,121 false positives out of 10,242 suspicious issues) false positives reported by {\methodname}, where the four categories of false positives comprise 8\%, 19\%, 44\%, and 29\% of the false positives.
We will try to eliminate these false positives in the future by adopting better image inpainting techniques and POS Tagging tools, and improving the algorithm of object selection.

\subsection{Threats to Validity}

First, we only use 226 seeds to generate 9,250 image pairs for testing. 
This is because we need humans to review the results of suspicious issues.
Therefore, we cannot generate unlimited caption pairs for testing, which makes it impossible to verify the results.
The precision may be different if we generate more test cases. To mitigate this threat, we generate a decent amount of test cases for each of the five IC systems, achieving similar precision.
In addition, due to the limitation of XOPS implemented by Stanza~\cite{stanza}, {\methodname} could report false positives.
In the future, we will try different tools and maintain a list of candidates that existing tools fail.
Another threat is that we only evaluate {\methodname} on five IC systems.
To reduce this threat, we select the most representative IC systems: one of these systems is a commercial API provided by Microsoft, and the other four are SOTA IC algorithms published in recent papers.





\section{Related Work}

\subsection{Robustness of AI Software}
The success of deep learning makes AI software the leading solution in many fields, but the parameter-based machine learning models that power these intelligent applications
are prone to failure.
The inferior results caused by the glitches of AI software may incur dissatisfaction or even severe accidents.
For example, Google apologized for Photo App's racist blunder that mistakenly labeled a black couple as being "gorillas"~\cite{racist_google}.
The traffic accidents~\cite{accident1, accident2, accident3, accident4} caused by autopilot systems severely threaten the safety of the users, which raised public concern about AI auto-driving techniques and slowed down their commercial deployment.
The research community has put much effort into researching the robustness of AI software~\cite{sun2018concolic, sun2019structural, sun2018testing, panichella2022test, zampetti2022using, khatiri2021machine, cao2022deepfd, wang2018object}.
Some proposed a variety of methods to attack the AI systems~\cite{athalye2018obfuscated, carlini2016hidden, carlini2017towards, liu2017trojaning, tao2018attacks, xu2020towards}, others try to design robust models~\cite{zhang2019towards, babenko2010robust, zhong2012robust, cheng2018towards} for AI software to cope with the intractable cases in the real world.

To enhance the robustness of AI software, testing techniques for AI software are flourishing~\cite{zhang2020machine, papadakis2019mutation, chen2022maat, xie2019deephunter, ma2018deepgauge,  humbatova2021deepcrime, pham2021deviate, ji2022asrtest, pan2022test, liu2022qatest, cao2022semmt}.
In contrast with many of the current methods that require the parameters of the target models, our method is totally black-box, which can be more conveniently adopted to test various IC systems without requiring any internal details.

\subsection{Robustness of CV Algorithms}
The AI-driven CV software brings both convenience and prospective danger to our daily life.
There are some examples of the danger: (1) criminals who use the photos to cheat the face recognition system, (2) autopilot systems fail to recognize the potential danger in front of the car.
There are several automated testing frameworks developed to test the robustness of CV algorithms.
DeepTest~\cite{tian_deeptest_2018} is a systematic testing tool for automatically detecting erroneous behaviors of autopilot through several image transformations.
MetaIC~\cite{yu2022automated}, the first work for testing automated IC systems, inserts an object into the background image and control the overlapping ratio of the inserted object and the original ones.
Despite high precision, MetaIC cannot tackle object classes with synonyms or word groups, while {\methodname} can cope with these cases.
Besides, the synthetic test cases generated by the current methods~\cite{tian_deeptest_2018, wang2020metamorphic, yu2022automated} barely consider the naturalness of the images.
{\methodname} can generate images that are more natural than the SOTA methods, which better simulate real-world scenarios.


\subsection{Metamorphic Testing}
Metamorphic testing~\cite{chen2020metamorphic} is an approach to both test case generation and test result verification, serving as an important technique to tackle the test oracle problem.
Over the years, metamorphic testing techniques have developed considerably in various perspectives, including metamorphic relation identification, test case generation, integration with other software engineering techniques, and verification of software systems~\cite{chen2018metamorphic}.
There are several automated testing tools adopting metamorphic testing in CV~\cite{tian_deeptest_2018, wang2020metamorphic, yu2022automated} and NLP~\cite{he2020structure, he2021testing}.
In this paper, {\methodname} uses an image inpainting technique to generate test cases, and identify several metamorphic relations to verify the results.

\section{Conclusion}\label{sec:con}

We propose {\methodname}, a novel, widely-applicable metamorphic testing methodology for image captioning systems.
Different from existing approaches that insert an object into the seed image, {\methodname} melts an object and assumes that the object set of the descendant is the subset of that of the ancestor.
Based on this core idea, we develop two MRs: object melt and class melt.
In our human study, the test cases generated by {\methodname} largely outperforms the test cases generated by the SOTA approach in terms of naturalness. 
Moreover, the test cases generated by our approach achieved comparable naturalness to the original images, demonstrating its superiority.
Meanwhile, {\methodname} successfully report 9,121 erroneous issues in one commercial IC API and four IC algorithms with high precision (86.47\%-92.17\%), which is comparable to the SOTA.
The test cases can be leveraged to fine-tune IC models, which effectively improves their performance. 
{\methodname} can also be adapted to clean the captioning dataset by unveiling labeling errors in COCO Caption.
We believe that the core idea of {\methodname} could be adapted to test various AI software (\textit{e.g.}, object detection software) in the future.

\section{Data Availability}\label{sec:data}

Codes and data of {\methodname} can be found at~\cite{romeCode}.

\begin{acks}
We thanks the anonymous ISSTA reviewers and Qiuyang Mang for their valuable feedback on the earlier draft of this paper. This paper was supported by the National Natural Science Foundation of China (No. 62102340), and Shenzhen Science and Technology Program.
\end{acks}

\balance

\bibliographystyle{ACM-Reference-Format}
\bibliography{References}


\end{document}